# Story and essential meaning dynamics in Bangladesh's July 2024 Student-People's Uprising

Tabia Tanzin Prama,[1,*] Christopher M. Danforth,[1,2,†] and Peter Sheridan Dodds[1,3,4,‡]

[1] *Computational Story Lab, Vermont Complex Systems Institute, Vermont Advanced Computing Center, University of Vermont, Burlington, VT 05405, US.*

[2] *Department of Mathematics and Statistics, Burlington, VT 05405, US.*

[3] *Department of Computer Science, Burlington, VT 05405, US.*

[4] *Santa Fe Institute, 1399 Hyde Park Rd, Santa Fe, NM 87501, US*

News media serves a crucial role in disseminating information and shaping public perception, especially during periods of political unrest. Using over 50,0000 YouTube comments on news coverage from July 16 to August 6, 2024, we investigate the emotional dynamics and evolving discourse of public perception during the July 2024 Student-People's Uprising in Bangladesh. Through integrated analyses of sentiment, emotion, topic, lexical discourse, timeline progression, sentiment shifts, and allotaxonometry, we show how negative sentiment dominated during the movement. We find a negative correlation between comment happiness and number of protest deaths ($r = -0.45$, $p = 0.00$). Using an ousiometer to measure essential meaning, we find public responses reflect a landscape of power, aggression, and danger, alongside persistent expressions of hope, moral conviction, and empowerment through goodnesses. Topic discourse progressed during the movement, with peaks in 'Political Conflict', 'Media Flow', and 'Student Violence' during crisis surges, while topics like 'Social Resistance' and 'Digital Movement' persisted amid repression. Sentiment shifts reveal that after the second internet blackout, average happiness increased, driven by the more frequent use of positive words such as 'victory', 'peace' and 'freedom' and a decrease in negative terms such as 'death' and 'lies'. Finally, through allotaxonometric analysis, we observe a clear shift from protest to justice.

## I. Introduction

The July Revolution, also known as the Student-People's Uprising, was a major pro-democracy movement that took place in Bangladesh in 2024 [1–3]. The uprising initially began as a quota reform movement in June, led by the group Students Against Discrimination [4, 5], following the Supreme Court's invalidation of a controversial government circular regarding public sector job quotas. What began as a peaceful protest quickly escalated into a nationwide uprising in response to violent government crackdowns, most notably the July massacre [6]. By August, the movement had transformed into a widespread non-cooperation campaign that ultimately resulted in the resignation of Prime Minister Sheikh Hasina. Her ousting triggered a constitutional crisis, leading to the formation of an interim government headed by Nobel Laureate Muhammad Yunus [7]. The uprising caused significant casualties: While initial government reports claimed 215 deaths [8], a UN investigation later confirmed that over 650 people were killed [9].

The period in question began when a reform movement gained momentum after the High Court upheld the quota system, and tensions escalated following the death of Abu Sayed in Rangpur on 16 July [10]. In response, the government imposed a nationwide internet shutdown on July 18 to curb online activism. Although partial internet access was restored on July 23 and fully restored by July 28, access to social media platforms remained blocked [11]. A second blackout occurred on August 2 amid renewed protests [12], and a violent crackdown on August 4 resulted in 91 additional deaths, prompting another shutdown. Internet services were restored after the fall of the government on August 5, although social media restrictions persisted [13]. Investigations later revealed that the former ICT Minister had directed the Bangladesh Telecommunication Regulatory Commission to disable internet services without formal approval [14].

The uprising evolved beyond quota reform, encompassing broader demands for democracy, government accountability, and freedom of expression. The human cost of the movement was profound, with death toll estimates ranging from several hundred to over 1,400. Shohid24[1] [15] is a publicly available open-source repository that documents the number of deaths during the July student movement, with a confirmed total of 456 deaths (accessed May 5, 2025). The dataset includes each victim's name, demographic information, biography, affiliations, verification sources, and details on how they were killed during the movement. While numerous studies have documented the deaths and offered historical accounts, no prior research has examined public perception of the uprising through the lens of mainstream news media coverage. We fill that gap by combining integrating sentiment, emotion, discourse, and timeline analyses together to explore how public perception changed over time and contribute to broader discussions on social justice, media framing, and protest movements in authoritarian and semi-authoritarian regimes [16–18]. While previous sentiment analysis studies on public perceptions of the economy, job security, and government responses have relied mainly on social media data, we focus here on public perception through comments on the news media due to repeated internet blackouts and social media blockages during the movement.

* tabia.prama@uvm.edu
† chris.danforth@uvm.edu
‡ peter.dodds@uvm.edu

## II. Description of datasets

### A. Data collection

To examine shifts in public perception from June 15 to August 5, 2024, we select 1000 protest-related news YouTube videos from Jamuna Television is also known as Jamuna TV[2]. Jamuna TV, a news organization based in Bangladesh, uses its YouTube channel to provide in-depth coverage of both domestic and international affairs. According to Social Blade rankings[3], it has recorded 24.1 billion views and 27.7 million subscribers, ranking 10th globally among the Top 50 News & Politics YouTube channels and 1st in Bangladesh (accessed May 5, 2025). From over 3,000 available videos on this movement, we filtered for those with more than 100,000 views and at least 200 comments to ensure content quality and audience engagement. In total, we collected approximately 50,000 public comments posted during the movement, which occurred amid two nationwide internet shutdowns and multiple social media blackouts. Comments were retrieved using the YouTube Data API. In addition, we used Shohid24 [15], an open-source dataset that documents the details of deaths of the July Student Movement.

[1] https://github.com/Shohid24/shohid24
[2] https://www.jamuna.tv
[3] https://socialblade.com/youtube/lists/top/50/sb/news/global

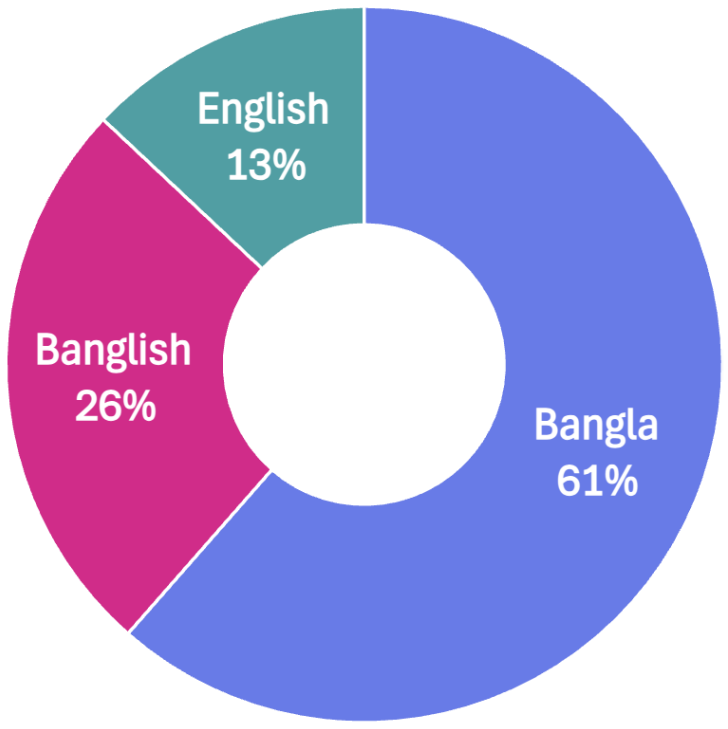

Figure 1. Distribution of languages in YouTube comments collected from videos related to the July movement.

### B. Text processing

After collecting comments from news reports, we manually removed promotional comments and those that contained only emojis. The remaining comments written in Bangla and Banglish (Bangla written in English) were then translated into English using the Google Cloud Translation API[4] to facilitate easier analysis. Figure 1 illustrates the language distribution of the comments. Appendix A1 Table A1 provides examples of the collected comments. We verified the accuracy of these translations through manual annotation and replaced emojis with corresponding emotion words using a demojization process[5] to support sentiment and emotion analysis.

For text preprocessing, we utilized Python, which included normalization (converting text to lowercase) and the removal of unnecessary elements such as duplicate entries, URLs, and non-English characters.

## III. Analysis and Discussion

### A. Correlation between public engagement and death toll

Figure 2 presents the number of comments posted on YouTube news reports during the movement period, while Figure 4 depicts the daily number of reported deaths based on the Shohid24 dataset. Comment activity spiked on March 16, 2024, following violent attacks by ruling party activists on quota protesters, which left more than 300 people injured and at least six dead. Engagement with news reports on that day tripled compared to the previous day, as nearly 10,000 students joined protests across the country.

[4] https://cloud.google.com/translate
[5] https://pypi.org/project/demoji/

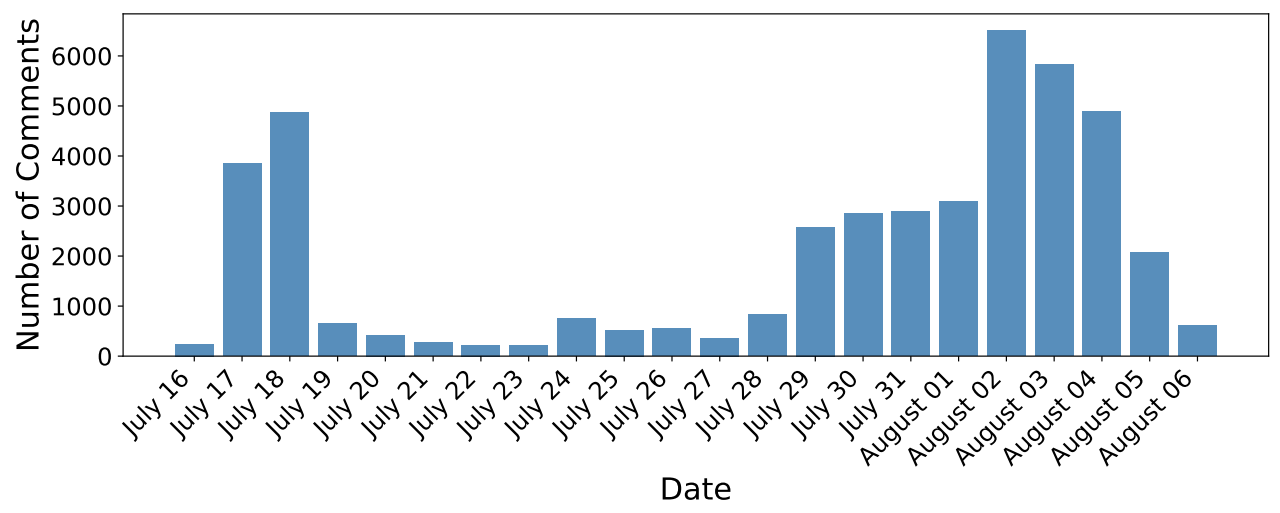

Figure 2. Number of comments on YouTube news reports related to the July Movement during the protest period. The bars represent the daily number of comments on July Movement-related news reports published on YouTube between July 15 and August 10, 2024. The y-axis indicates the number of comments, while the x-axis shows the timeline.

Further surges in violence occurred on July 17 and 18, when coordinated assaults by police and members of the Bangladesh Chhatra League (BCL)—the student wing of ousted Prime Minister Sheikh Hasina's Awami League party—targeted student protesters on campuses across 19 districts. These attacks resulted in at least 150 deaths. According to the United Nations report on the protests of July and August 2024 in Bangladesh [19], the Ministry of Health recorded 841 deaths linked to the protests, including 10 women, as well as 12,272 injuries, among them 394 suffered by women and four by individuals listed as "other." Figure 3 illustrates the nationwide distribution of deaths and injuries based on Ministry of Health data [20].

Amidst this violence, engagement on movement-related news videos surged. Comment activity peaked just before the first internet blackout (July 18–28) and again prior to the second blackout (August 4–5), both of which were followed by significant increases in the death toll. During the blackouts, comment volume dropped notably, as people within Bangladesh were unable to access YouTube. However, some engagement continued from users located abroad. Using Pearson correlation analysis, we reveal a statistically significant positive correlation between the number of comments and the number of reported deaths during the movement period ($r = 0.368, p = 0.02 < 0.05$), suggesting a connection between rising public engagement and the number of deaths.

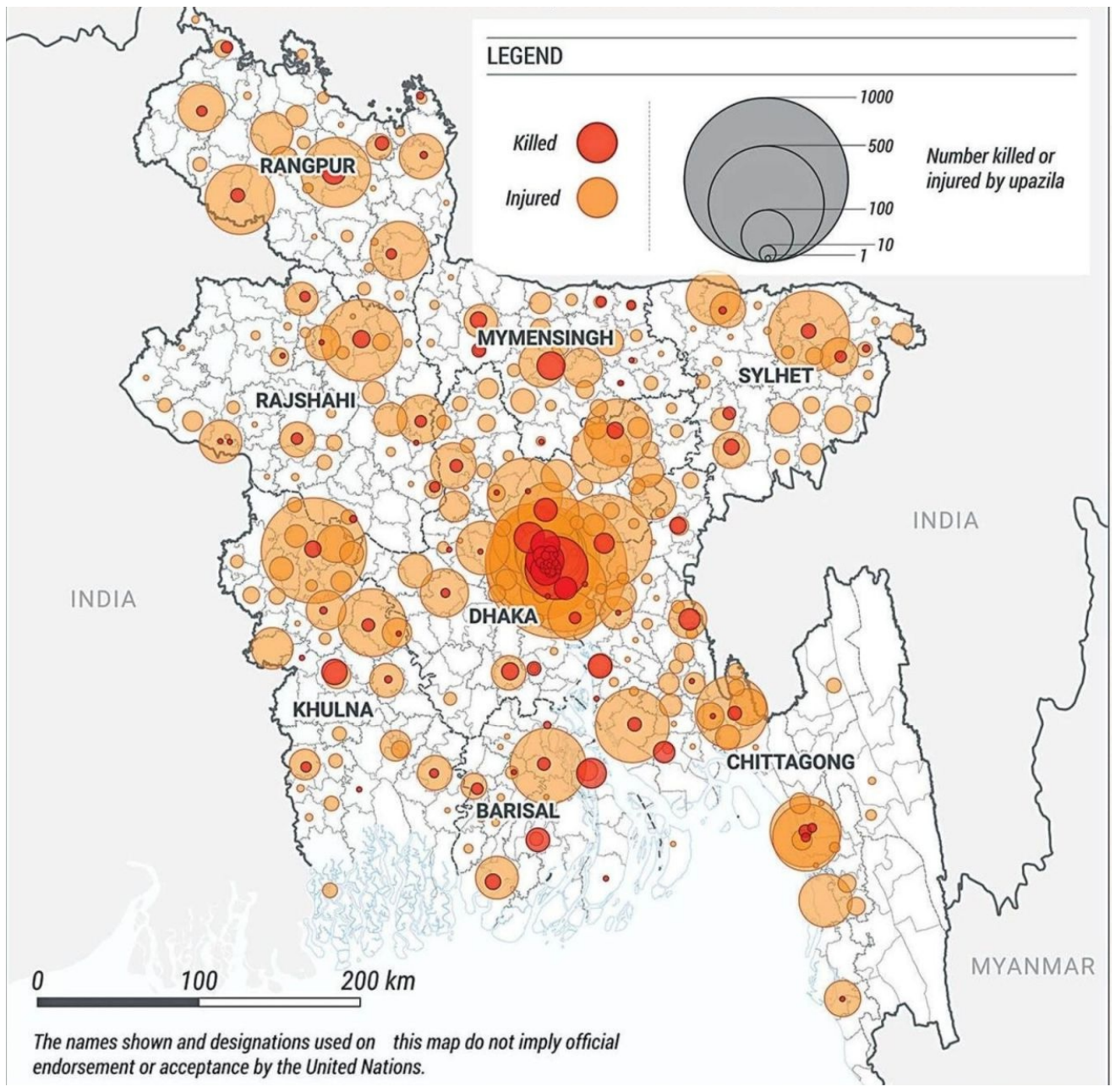

Figure 3. Distribution of Deaths and Injuries Recorded by the Ministry of Health in the Context of Protests. (Source: United Nations Human Rights Office (OHCHR)) [19].

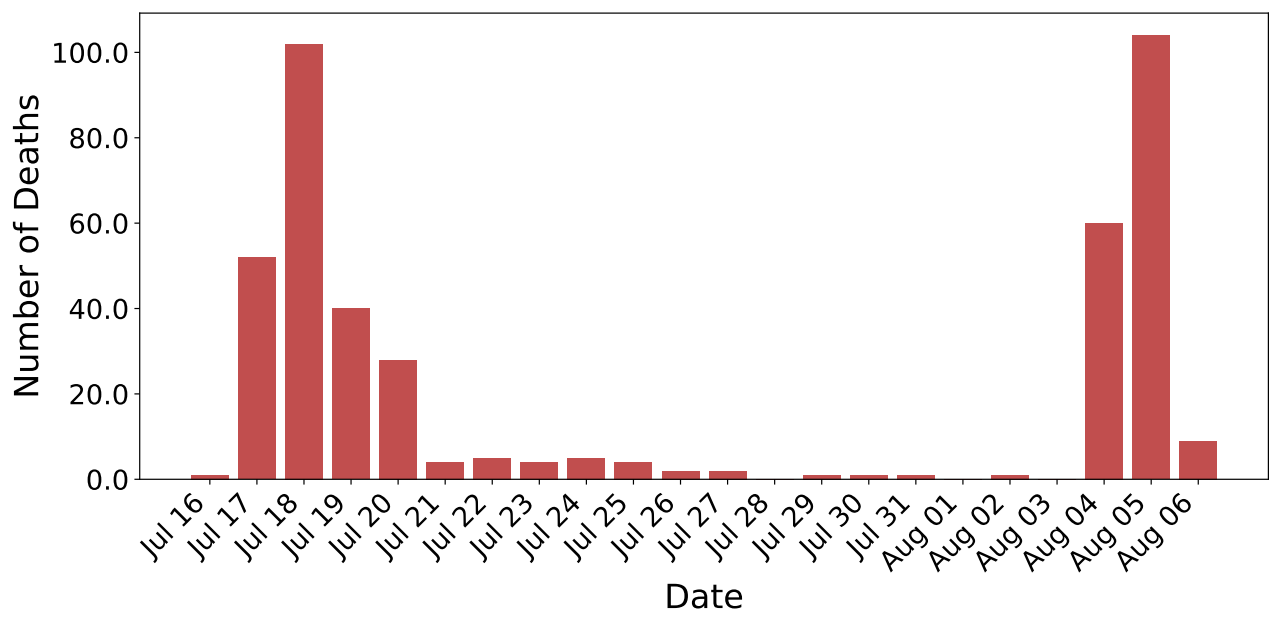

Figure 4. Daily number of reported deaths during the July Student Movement. The bars represent the number of deaths reported each day during the July Student Movement, based on the Shohid24 [15] dataset.

## B. Extraction of sentiment

To measure societal sentiment during the movement, we calculated the average happiness score for public comments using a text-based hedonometer [21]. The hedonometer[6] is based on a list of English words annotated with average happiness scores ranging from 1 to 9, representing a spectrum from sad to happy ($h_{\text{avg}} = 5$ indicates neutrality, $h_{\text{avg}} > 5$ corresponds to positive emotions, and $h_{\text{avg}} < 5$ reflects negative emotions). We calculated ($h_{\text{avg}}$) for each comment independently on a daily basis, allowing us to track the evolution of sentiment throughout the different stages of the movement.

[6] http://hedonometer.org

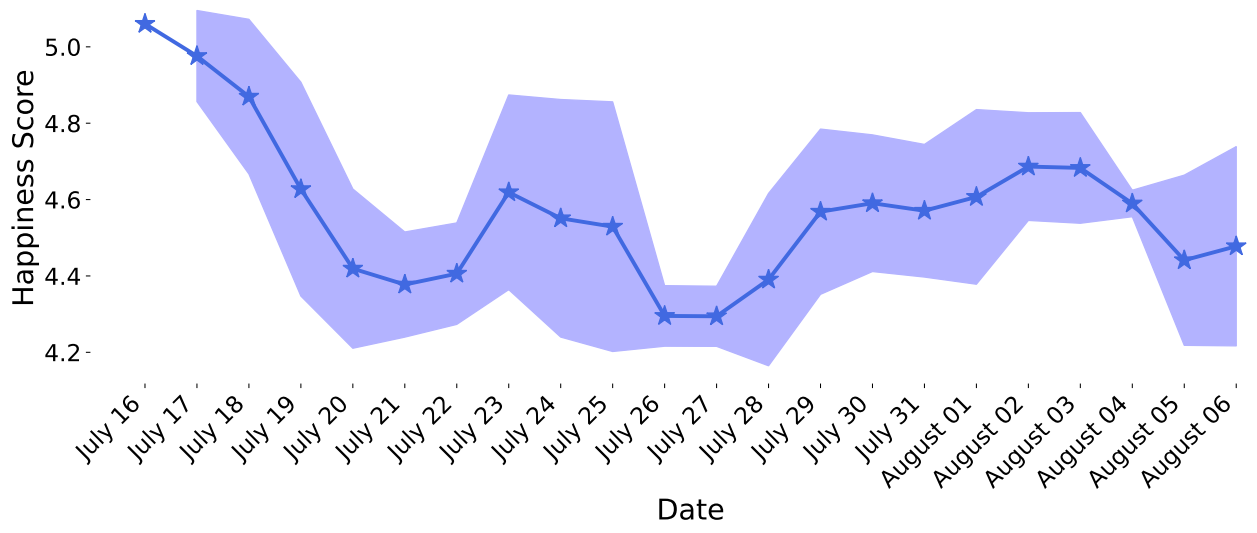

Figure 5. The daily average happiness score using the Hedonometer from July 16 to August 6, 2024, covering the movement period.

Figure 5 shows the trajectory of the average happiness score of public comments on movement-related news reports during the protest period. Throughout the entire time span, we observed substantial, system-wide declines in happiness in response to various events, both exogenous and endogenous in nature, such as large numbers of fatalities and internet blackouts. At the beginning of the time series, we notice a sharp decline in the happiness score following the massive protests on July 17, after several protesters were killed. The government's announcement of a nationwide curfew induced a multi-day drop in happiness through July 21. During the first phase of the internet blackout, the happiness score fluctuated as the curfew was extended indefinitely, and key organizers of the quota demonstration were detained. After the first internet blackout concluded on July 28, some vowed to continue the protests which erupted again in multiple districts, leading to further police clashes and detentions. The average happiness score also rose slightly, reflecting public activation. Similarly, prior to the second internet blackout (August 4–5, 2024), massive killings of protesters resulted in another large drop in the average happiness score, reflecting the public's distress. There is a clear relationship between the number of deaths and the average happiness score, with both changing in tandem as the movement progressed. Generally, following incidents of mass killings, the average happiness score drops drastically[7]. A Pearson's correlation analysis reveals a significant negative correlation between the daily average happiness score and the number of deaths during the July Student Movement ($r = -0.45, p = 0.00, p < 0.05$).

[7] http://hedonometer.org

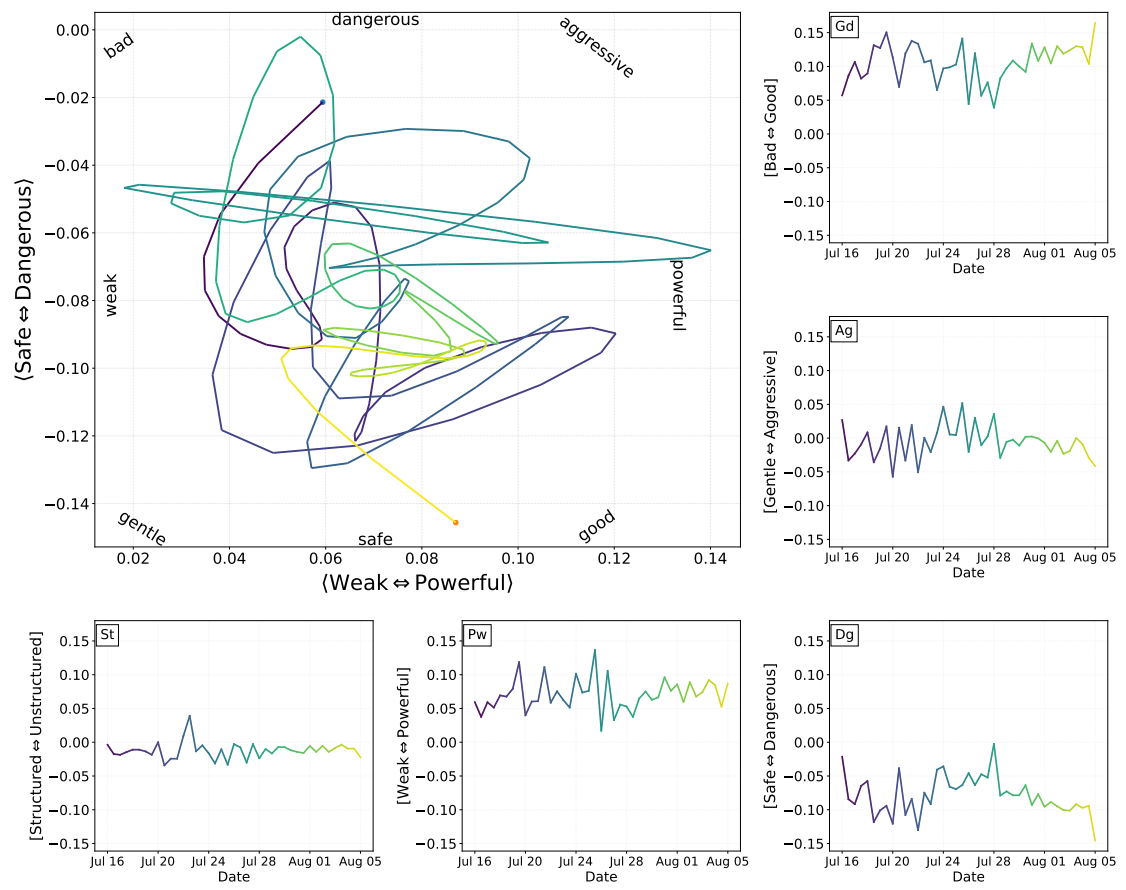

Figure 6. Public emotional dynamics on YouTube during the movement period, as measured by the Ousiometer. The main trajectory plot (left) shows the emotional changes over time within the DAPGS (Danger-Aggression-Power-Goodness-Structure) framework. The time series starts in blue and ends in red. The five subfigures represent the change in each dimension of the DAPGS framework throughout the movement period: Top right (Gd: Bad ⇔ Good), Middle right (Ag: Gentle ⇔ Aggressive), Bottom right (Dg: Safe ⇔ Dangerous), Bottom left (St: Structured ⇔ Unstructured), and Bottom middle (Pw: Weak ⇔ Powerful). Each value in the time series is calculated as the mean of the dimension value for that day. The color of the line transitioning from purple to yellow as the movement progresses from start to end (July 16th to August 6).

### C. Emotion Analysis

To gain deeper insights into the emotional undercurrents that shaped public discourse during the Bangladesh July 2024 Movement, we employed an advanced tool known as the ousiometer, a lexical instrument designed to quantify emotional and semantic tones within large-scale textual data [22]. Building on the principles established by the hedonometer, the ousiometer is capable of measuring a wide range of emotional responses through DAPGS (Danger-Aggression-Power-Goodness-Structure) framework framework. These framework allow us to understand the multidimensional emotional landscape of the public comments during the protest period.

We processed comments from YouTube, filtering them to retain only the 1-grams (individual words) that appeared in the corresponding ousiometric lexicons. For each word $\tau$ in the filtered text, we computed its normalized frequency $p_\tau$ as:

$$p_\tau = \frac{f_\tau}{\sum_{\tau'} f_{\tau'}}$$

where $f_\tau$ is the frequency of word $\tau$, and $\sum_{\tau'} f_{\tau'}$ is the total frequency of all words in the text. Using this normalized frequency, we then calculated the average emotional score for each comment as:

$$M_{\text{avg}}(\Omega; L) = \sum_{\tau \in R_L(\Omega)} p_\tau \cdot M_\tau$$

where $M_\tau$ is the emotional score associated with each word in the lexicon. The resulting scores were averaged daily over the course of the movement, generating the time series shown in Figure 6, which presents the evolution of public emotions across the movement period from July 16 to August 6, 2024.

The main trajectory plot in Figure 6 illustrates the fluctuations in public emotional states in response to key events throughout the protest period. The results reveal a striking pattern in the emotional dynamics, with distinct phases corresponding to major turning points in the movement. Initially, the emotional landscape was dominated by fear and anger, reflecting the public's response to mass fatalities and government repression. As the protests continued, there was a shift toward empowerment and resilience, culminating in a surge of positive emotions following the resignation of the government.

### *1. Early Stages: Rising Danger and Aggression (July 16–19, 2024)*

The first few days of the movement were marked by escalating violence, particularly due to police killings of student protesters. The emotional response during this period was characterized by heightened danger and aggression, as reflected in the rise of the Danger (Dg) and Aggression (Ag) dimensions of the emotional dynamics. As shown in Figure 6, the Danger dimension spiked as the protests became increasingly violent and people feared for their safety. At the same time, the Aggression dimension showed a notable increase, reflecting the public's rising anger and frustration in response to government repression.

This emotional surge was a direct consequence of the mass killings that occurred on July 17, when several protesters were killed by police forces. The movement's escalation, coupled with the government's refusal to address the demands of protesters, amplified the public's sense of danger and the collective feeling of aggression. During this period, emotions were raw, and online discourse reflected a sense of urgency and resistance, captured by the high values in the Aggression dimension, as citizens voiced their anger through social media.

### *2. The Midpoint: Transition Toward Power and Resilience (July 20–28, 2024)*

As the movement progressed, particularly during the first phase of the internet blackout, the emotional dynamics shifted. The Danger dimension began to decline as the immediate physical threat from police receded, but the Aggression dimension remained elevated, signaling ongoing frustration and resistance. The Power (Pw) dimension, which measures empowerment and collective agency, saw a gradual increase as the protests spread and more people joined in defiance of the government.

The period from July 20 to July 28 saw a marked shift in emotional tone. As the protestors defied curfews and continued their demonstrations in the face of extreme repression, the Power dimension surged, reflecting the growing sense of collective strength. During this time, there was also a noticeable increase in structure as the protests gained organization, particularly in the face of digital repression and internet censorship. The Structure (St) dimension rose during the internet blackout, indicating a more unstructured, chaotic, and disorganized nature of communication as people struggled to coordinate their efforts.

### *3. Climax: A Surge in Power and Goodness (July 29–August 5, 2024)*

The emotional landscape took another significant turn following the second wave of mass killings prior to the second internet blackout on August 4-5. The Danger and Aggression dimensions again spiked, reflecting public outrage over the loss of life and the continued violence against protesters. However, following the government's eventual resignation on August 5, a dramatic shift occurred. The Goodness (Gd) dimension, which had been low throughout the earlier violent stages of the protest, surged to its highest point, reflecting the public's sense of relief, hope, and victory after the government's collapse.

Simultaneously, the Power dimension remained high, as the protestors celebrated their collective success and the achievement of their goals. This period marked a sense of empowerment, with thousands of people celebrating in the streets. The sharp increase in Goodness reflected the collective emotional relief and triumph after the protests had achieved their primary objective—bringing down the government.

*4. Final Stages: Celebration and Reflection (August 6, 2024)*

The final phase of the movement was characterized by a shift in emotional tone from aggression and fear to celebration and triumph. As shown in the time series, the Goodness dimension reached its peak, as people expressed a sense of victory, moral affirmation, and hope for the future. The Danger dimension continued to decline, reflecting the decreasing immediate threat to public safety, while Aggression also reduced, signaling a shift from confrontation to celebration.

| Emotion | Mean | Standard Deviation |
|---|---|---|
| Goodness | 0.11 | 0.17 |
| Aggression | -0.01 | 0.13 |
| Structure | -0.01 | 0.06 |
| Power | 0.07 | 0.15 |
| Danger | -0.09 | 0.16 |

Table I. Descriptive statistics for the five emotional dimensions in DAPGS (Danger-Aggression-Power-Goodness-Structure) Emotion Frameworks for dataset.

Table I shows the mean and standard deviation of eight emotional dimensions extracted from YouTube comments. Negative mean values for Danger (-0.09), and Aggression (-0.01) indicate an emotional climate of suppression, low energy, and heightened threat. Positive mean values for Goodness (0.11), Dominance (0.06), Power (0.07), suggest optimism, moral support, and hope. The high standard deviation of Goodness (0.17) and Danger (0.16) reflects emotional volatility during critical moments such as violent escalations or internet blackouts. This analysis reveals the emotional landscape of the protest, characterized by both optimism and resistance, alongside the challenges of danger and aggression.

Also, Table II shows the most frequent words categorized by emotional dimensions from YouTube comments during the 27-day protest. Words like 'winning,' 'freedom,' 'victorious," and 'wisdom' fall under Goodness, highlighting themes of hope, moral affirmation, and collective agency. In contrast, Danger-related terms such as 'murderer,' 'terrorism,' 'killer,' and 'slaughter' reflect the heightened threat and risk faced during the protest. Power-associated words like 'almighty,' 'victory,' and 'champion' suggest empowerment, while Aggression-related terms such as 'missiles,' 'battle,' and 'killing' emphasize the conflictual and violent aspects of the movement. These words effectively capture the emotional intensity and complexity of the protest.

| Goodness | Aggression | Structure | Power | Danger |
|---|---|---|---|---|
| confidence | missiles | play | success | murderer |
| perfect | battle | laughs | almighty | murderous |
| freedom | firing | joking | awesome | killer |
| wining | kill | party | victorious | dangerous |
| wisdom | terrorists | running | champion | terrorist |
| reliable | battlefield | impressed | powerful | terrorism |
| honorable | terrorism | surprise | winning | terrorists |
| briliant | killing | dance | successful | slaughter |
| success | sniper | popcorn | victory | kill |
| excellent | guerrilla | joker | succeed | hell |

Table II. Top 10 Words Based on the DAPGS (Danger-Aggression-Power-Goodness-Structure) Emotion Frameworks from YouTube Comments During the July Movement

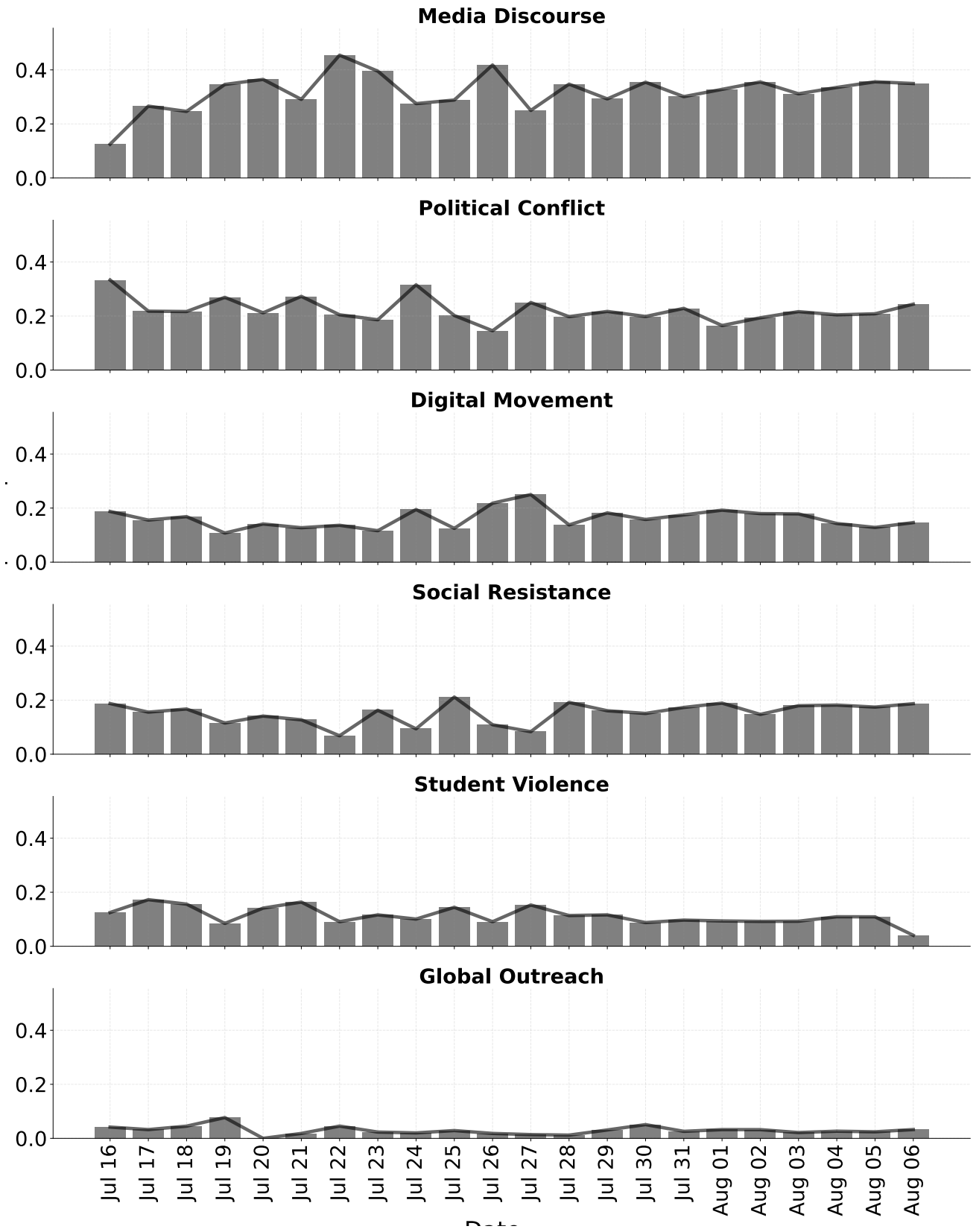

Figure 7. Topic distribution of public perception based on YouTube data from July 15 to August 6 (from top to bottom: most to least prominent topics).

## D. Topic Analysis

To conduct topic analysis, we applied Latent Dirichlet Allocation (LDA) [23] and selected 10 topics. Appendix A2, Table A2, presents the topic labels along with their associated words. Since several topics were closely related, we merged them into broader categories. Table A3 shows the final set of merged topics. Figure 7 il-lustrates the temporal dynamics of these topics during the movement period. Among them, media discourse and political conflict emerged as the two most dominant themes. Discussions under the media discourse category increased significantly during the first week and remained highly prominent throughout the movement, while political conflict reflected the public's ongoing concerns with political leadership and regime accountability. By contrast, digital movement and social resistance were less prominent at the outset but grew steadily from the middle of the movement onward. These themes intensified in parallel with escalating violence, rising casualty numbers, and periods of restricted internet access, often spiking immediately after major incidents. Student violence was a dominant theme in the early phase, particularly between July 17–19, following the first wave of student killings. Meanwhile, global outreach remained comparatively limited but showed small peaks between August 1–5, suggesting attempts to attract international attention during the final week of the movement.

### E. Sentiment Dynamics of Internet Blackouts During the Movement

Dictionary-based sentiment analysis is inherently sensitive to the dictionary used. Sentiment dictionaries are often static, constructed once for general use. This approach can be problematic when there has been a temporal shift in the meaning of particular words, or when words take on different sentiments in specific contexts [25]. Word shift graphs [26] transparently diagnose these types of measurement issues.

To visualize how sentiment shifts over the course of the movement and provide a meaningful, interpretable summary of how individual words contribute to the variation between different phases, we divided the entire movement period into four phases: before the first internet blackout (I.B) (15th July - 18th July), during the internet blackout (19th July - 28th July), before the second internet blackout (29th July - 3rd August), and during the second internet blackout (4th August - 5th August). We applied the labMT sentiment dictionary [27] to public comments on news reports from the periods before and during the internet blackouts. This simple quantification offers an emotional arc of the public's sentiment during the movement. For both word shift graphs in Figure 8, a reference value of 5 was used, and a stop lens was applied to all words with a sentiment score between 4 and 6. Both word shift graphs include cumulative contribution and text size diagnostic plots in the bottom left and right corners, respectively. We measured the difference in average sentiment between the movement phases: before and during the first I.B. (on the left panel), and before and during the second I.B. (on the right panel) of Figure 8, ranking words by their absolute contribution to the sentiment difference. First, examining the left panel of Figure 8 reveals that during the I.B., sentiment is higher due to a preponderance of less frequent negative words and more abundant positive words. The contribution of the former is marginally larger than that of the latter.

The word shift graph for the first internet blackout period shows that positive words ($+\uparrow$) such as 'people,' 'thanks,' 'justice,' 'love,' 'victory,' 'Understand,' 'Party,' 'Mother,' 'Beautiful,' and 'World' increased the average happiness ($\Phi_{\text{avg}}$) over the baseline. These positive shifts were accompanied by decreases in the prevalence of negative words ($-\uparrow$) such as 'died,' 'death,' 'arrested,' 'destroyed,' and 'punishments.' However, some positive words, such as 'students,' 'bothers,' 'university,' 'children,' 'right,' and 'education,' ($+\downarrow$) appeared less frequently, indicating a shift in public conversations from attacks on students to hope for victory in the movement. Overall, the largest contribution to the increase in $\Phi_{\text{avg}}$ of 0.09 came from the increase in positive words ($+\uparrow$) during the blackout period (18th July - 28th July).

The right panel of Figure 8 shows the word shift graph for the sentiment difference between the period before the second internet blackout (I.B.) and during the second I.B. We observe an increase in $\Phi_{\text{avg}}$ of 0.07, driven by relative increases in postive words ($+\uparrow$) such as 'party,' 'love,' 'life,' 'freedom,' 'thanks,' 'peace,' 'country,' 'victory,' 'power,' 'mother,' 'star,' 'grant,' 'real,' and 'done'. These terms reflect the celebration of the movement's victory. Simultaneously, there was a decrease in the occurrence of negative terms ($-\downarrow$) like 'killing,' 'blood,' 'death,' and 'lies.' Similar to the first phase of the internet blackout, the more frequent use of positive words ($+\uparrow$) and the less frequent use of negative words ($-\downarrow$) contributed substantially to the sharp increase in average happiness during the second internet blackout (4th August - 5th August).

### F. Allotaxonometry of Public Comments During the Movement

In Figure 9, we present an allotaxonomic *rank-rank histogram* with *rank-turbulence divergence* [28] to compare word usage across two key phases of the movement, as reflected in YouTube comments. The first phase spans from the start of the movement until the Internet Blackout (I.B) on 18th July 2024, and the second phase runs from the I.B on 28th July 2024 to the end of the movement on 5th August 2024. For this analysis, we first removed stopwords and emojis, then extracted words as *1-grams* (contiguous sequences of non-whitespace characters) from the translated comments. We then determined separate ranked lists of 1-grams for each phase

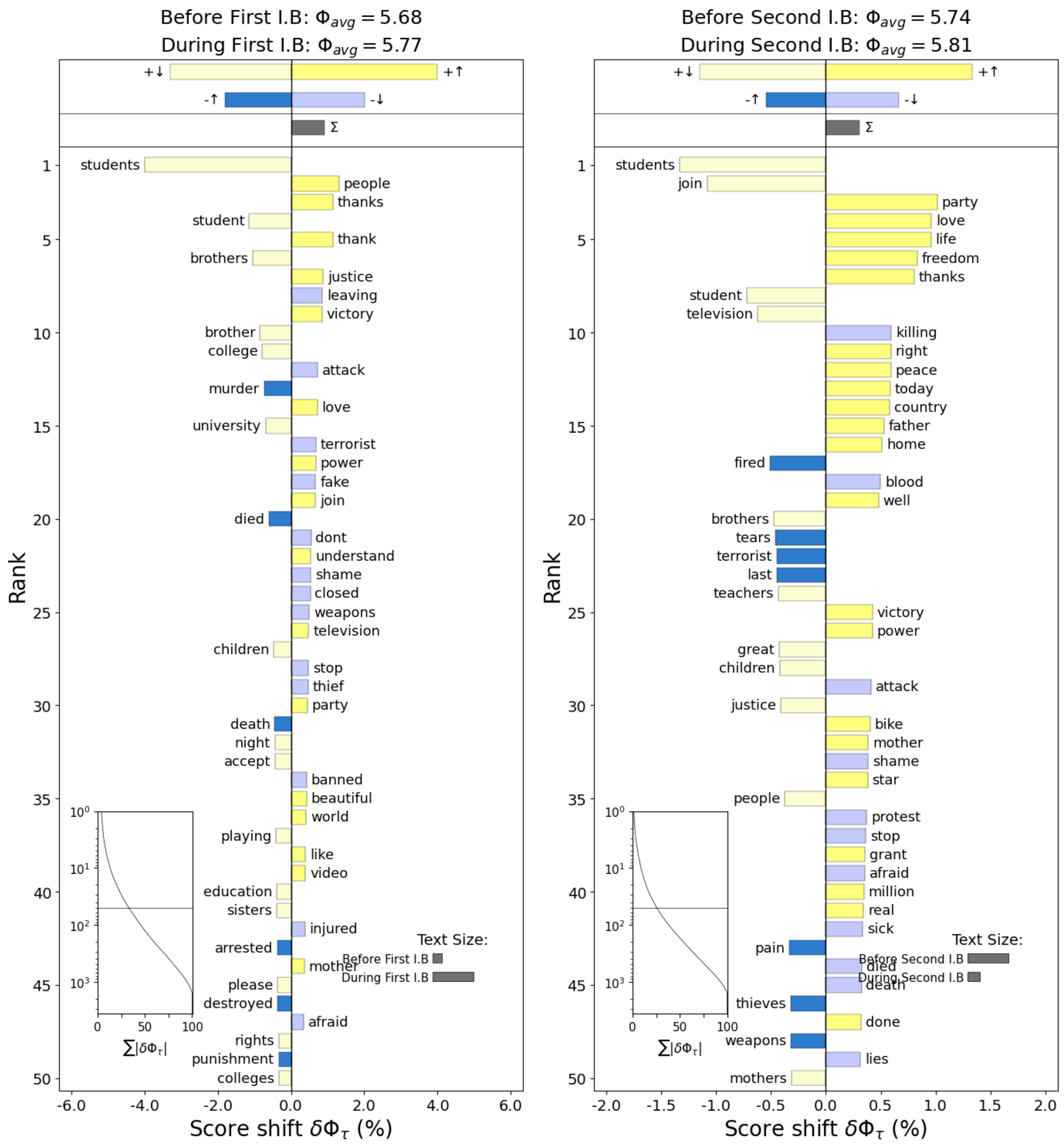

Figure 8. Word shift graph of word frequencies in happiness during two internet blackout periods of the July movement. Words are ranked by their percentage contribution to the change in average happiness, $\Phi_{\text{avg}}$. The days before the internet blackout (15th July - 18th July, left figure) and (29th July - 3rd August, right figure) are set as the reference text $T_{\text{ref}}$, with the respective internet blackout dates as the comparison text $T_{\text{comp}}$. Individual word contributions to the shift are indicated by two symbols: $+/-$ shows the word is more/less prevalent in $T_{\text{comp}}$ than in $T_{\text{ref}}$ . Black and gray fonts encode the $+$ and $-$ distinctions, respectively. The left inset panel shows how the ranked 3,686 labMT 1.0 words (Data Set S1) combine (word rank $r$ is shown on a log scale). The four bar on the top indicate the total contribution of the four types of words $(+\uparrow, +\downarrow, -\uparrow, -\downarrow)$. Relative text size is represented by the areas of the gray squares [24]

.

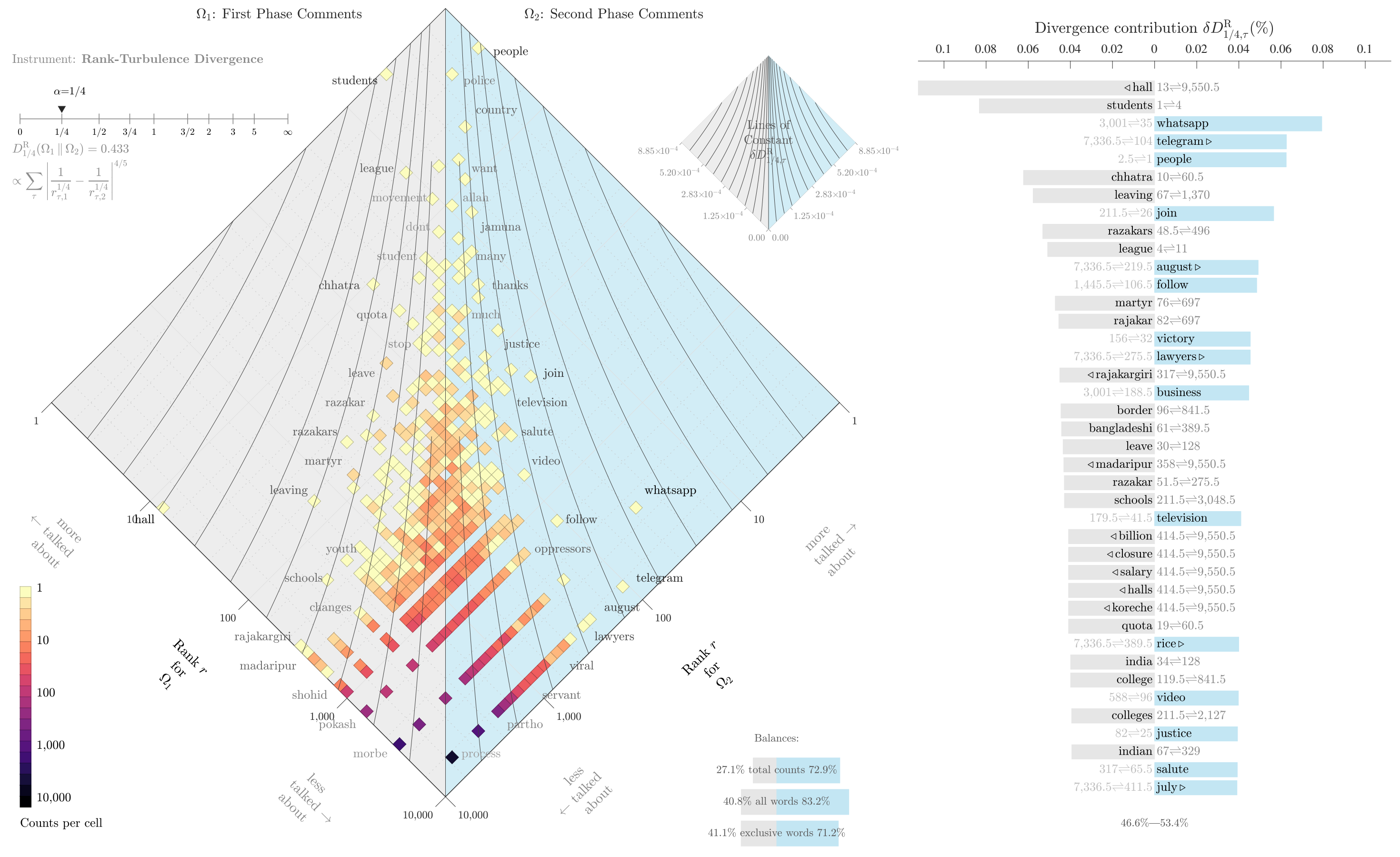

Figure 9. An example allotaxonograph using rank-turbulence divergence (RTD) to compare YouTube comments during two phases of a movement. The first phase spans from the start of the movement until the Internet Blackout on 18th July 2024, and the second phase runs from after the Internet Blackout on 28th July 2024 to the end of the movement on 5th August 2024. The left panel shows the rank-rank histogram, while the right panel shows the rank-turbulence divergence graph to visualize the differences in word usage between the two phases. The allotaxonograph compares ranked lists of types for two systems, $\omega_1$ and $\omega_2$, by first generating a merged list of types covering both systems and binning logarithmic rank-rank pairs $\log_{10} r_{\tau,1}, \log_{10} r_{\tau,2}$ across all types and uniformly in logarithmic space [28]. The discrete, separated lines of boxes nearest to each bottom axis comprise words that appear only in one phase: ‘exclusive types’. Moving up the histogram, two other distinct lines above the ‘exclusive-type lines’ correspond to words that appear once and twice in the other phase. The three horizontal bars in the lower right show system balances: the top bar indicates the balance of total counts of words (tokens) for each phase: 33% versus 67%. The middle bar shows the percentage of the combined lexicon (types) for the two phases that appear in each phase: 41% versus 83%. The bottom bar shows the percentage of words (types) on each phase that are exclusive: 42% and 71%. The rank-turbulence divergence graph on the right is based on contributions of each word to the divergence measure. An ordered list is presented by descending values of divergence. Words are arranged left and right and colored gray and blue based on which phase they are more prevalent in. Ranks for each word in both phases are shown: for example, $r_{\text{Hall},1} = 14$ and $r_{\text{Hall},2} = 9611$. The allotaxonograph limits the resolution of the divergence measure $\alpha$ to multiples of 1/12; here, $\alpha = 1/3$.

based on their frequencies. The Allotaxonometer compares the ranked lists of types for two systems (i.e., two phases of the movement), first generating a merged list of types covering both systems and then bin the logarithmic rank-rank pairs ($\log_{10} r_1, \log_{10} r_2$) across all types and uniformly distribute them in logarithmic space. Bins on either side of the central vertical line represent words that are more frequently used in the corresponding phase.

For example, 'Rajakar' (a Bangla term for 'traitor,' referring to individuals who opposed Bangladeshi independence and collaborated with the Pakistani Army) was ranked 51st before the Internet Blackout and 1179th afterward, while the term 'lawyers' moved from 7383rd before the blackout to 275th after. The discrete, separated lines closest to the bottom axes represent words that appear exclusively in one phase (i.e., exclusive types). Moving up the histogram, two other distinct lines correspond to words that appeared once or twice in the other phase. The three horizontal bars in the lower right of the histogram show system balances. The top bar indicates the balance of total word counts (tokens) for each phase: 33% before the Internet Blackout versus 40.1% after. The middle bar shows the percentage of the combined lexicon (types) that appears in each phase: 41% versus 83%. The bottom bar represents the percentage of words exclusive to each phase: 42% and 71%, respectively. Types appearing towards the top of the histogram rank high in both systems. For example, the 1-gram 'student' (movement also known as Student–People's uprising) is the most common word in both phases, with $r_{\mathrm{RT},1} = r_{\mathrm{RT},2} = 1$. The word 'student' is also significant, indicating that students were central to the movement from the start to the end. The words 'people' and 'country' are ranked 2nd and 3rd in both phases, while 'government' and 'league' (referring to *Bangladesh Chhatra League*, the student wing of the political party, often accused of committing violence) are ranked 4th and 5th before the Internet Blackout, but their positions reverse after the blackout. These high-ranking words reflect the topics most discussed and involved in the movement.

Moving down the histogram, turbulence begins to emerge around $r = 10$, with less common and more differentiating words appearing. Words that appear farthest from the central vertical axis show the greatest relative change in rank. Before the Internet Blackout, the word 'Hall' (referring to student dormitories or residence halls in public universities in Bangladesh, which became focal points for the movement) stands out as the movement originated from student dormitories. Further down, words like 'Razakars' (a derogatory term for 'traitors') and 'India' are prominent. After the blackout, terms like 'WhatsApp' and 'Telegram' gain prominence, as social media platforms like Facebook and Instagram were blocked by the government on 2nd August, prompting people to discuss alternative communication platforms.

While words like 'Hall' and 'WhatsApp' dominate the sides of the histogram, seemingly unrelated names and events also appear. On the right side, we see 'Jamuna' and 'Hasina'. Sheikh Hasina, the Prime Minister of Bangladesh during the movement, was blamed for the violence during the protests in July, as her government had clamped down on previous demonstrations. On the left side, we find 'school' and 'college', as students from these institutions were actively involved in the movement. The presence of the term 'India' signifies the geopolitical implications of the movement, especially its influence on regional connectivity and economic projects.

The right side of Figure 9 presents the *rank-turbulence divergence graph*, which provides a tunable single-parameter instrument for exploring the differences between the two systems by listing the types that contribute most to the divergence. We use rank-turbulence divergence with $\alpha = 1/3$. In this graph, 'Hall' and 'WhatsApp' stand out among the histogram annotations, while common words like 'students', 'country', and 'people' remain prominent, and rare words that appear only in one phase (like 'weather' and 'paltaia') are backgrounded. The rank-turbulence divergence graph orders the top 30 words by decreasing $\delta D R^{R}_{1/3,\tau}$, with words oriented left or right based on their higher rank in either phase. For example, 'Hall' has the highest divergence contribution, moving from rank 14 before the blackout to rank 9611 after. Conversely, 'WhatsApp' scores similarly to 'Hall' but shows a dramatic rank change, moving from 3021st to 36th.

The transition between the two phases is evident in the allotaxonomic graph, which reveals how the focus of the discourse shifted before and after the Internet Blackout. Before the blackout, the people comments were more focused on slogans and protesters, while after the blackout, the conversation shifted to justice, celebrating victories, and alternative social media platforms.

## IV. Concluding remarks

By integrating sentiment, emotion, lexical analysis, topic discourse, sentiment shifts, and allotaxonomy, this study offers a comprehensive view of public perception during the July 2024 Student–People's Uprising in Bangladesh using large-scale YouTube comment data. Sentiment analysis revealed consistent declines in public happiness during violent episodes—such as student killings and blackouts—signaling collective grief and outrage, with brief rebounds during periods of solidarity and hope. Emotion analysis showed a dual emotional landscape: fear and suppression reflected in negative arousal, energy, and danger scores, contrasted with persistent expressions of hope, moral conviction, and empowerment.

Topic modeling closely followed the protest timeline, with spikes in Political Power, Media Flow, and Justice Demand during key events. Sentiment shift analysis during blackouts showed rising happiness, driven by increased use of positive words like 'Victory' and 'Freedom,' and decreased use of negative terms. Allotaxonomic analysis captured a shift from protest-focused terms like 'Hall' and 'Rajakar' to post-blackout discourse emphasizing digital adaptation and resilience, with 'WhatsApp,' 'Telegram,' and celebratory language gaining prominence, while core terms like 'students,' 'people,' and 'country' remained central throughout.

Despite its contributions, this study has certain limitations. We analyzed comments from a selected set of 500 protest-related YouTube videos, and although they span major news coverage, they do not encompass the entirety of available content or represent all national TV channels in Bangladesh. Additionally, while social media platforms like Facebook and Twitter are significant venues for public discourse, data collection restrictions led us to focus solely on YouTube comments. The study also relied on translated Bangla comments and lexicon-based sentiment and emotion scoring (e.g., labMT and ousiometer lexicons), which may not fully capture contextual nuances. As Bangla is a low-resource language, word translations can be ambiguous and context-dependent, can potentially affect accuracy. Future research should prioritize the development of native sentiment lexicons for Bangla and leverage large language models (LLMs) to enhance contextual understanding and reasoning. Such approaches could enable broader analyses of public perception across global movements, from the Russia–Ukraine conflict to the Black Lives Matter protests [18, 29], offering deeper insights into digital civic engagement and emotional response during times of crisis.

## V. Data and Code Availability

All source code used for data processing, analysis, and visualization is available in this github repository 2024-Student-People's-Uprising.

## Acknowledgments

The authors are grateful for National Science Foundation Award #2242829 (Science of Online Corpora, Knowledge, and Stories), foundational support from MassMutual, and an anonymous philanthropic gift.

[1] The july revolution (bangladesh) (2024), accessed April 22, 2025.

[2] S. Moral, Student–people uprising: More than 18,000 injured (2024), published: 07 Sep 2024, 07:31.

[3] Staff Correspondent, 875 die in student–people uprising, most were bullet-hit: Hrss, accessed: May 25, 2025.

[4] 2024 bangladesh quota reform movement (2024), accessed: April 22, 2025.

[5] Students against discrimination (2024), accessed: April 22, 2025.

[6] S. C. M. Post, Bangladesh student protests become 'people's uprising' after brutal crackdown (2024), accessed: May 25, 2025.

[7] Interim government of muhammad yunus (2024), accessed: April 22, 2025.

[8] Bangladesh mourns some 200 deaths as student protests wind down and thousands are arrested (2024).

[9] Preliminary analysis of recent protests and unrest in bangladesh, *Country Reports* (2024), accessed: April 22, 2025.

[10] Abu sayed (student activist) (2024), accessed: April 22, 2025.

[11] A. Jazeera, Bangladesh restores mobile internet after 11-day blackout to quell protests (2024), ccessed: April 22, 2025.

[12] Facebook shut down again on mobile network, telegram too (2024), accessed: April 22, 2025.

[13] At least 91 killed in bangladesh protests as curfew and internet blocks imposed (2024), accessed: April 22, 2025.

[14] Internet was shutdown in bangladesh on 'verbal' orders from palak: probe report (2024), retrieved 24 October 2024.

[15] Shohid24 - a website listing the martyrs of the july student movement 2024 in bangladesh, https://shohid24.pages.dev/, accessed: May 5, 2025.

[16] R. J. Gallagher, A. J. Reagan, C. M. Danforth, and P. S. Dodds, Divergent discourse between protests and counter-protests: #blacklivesmatter and #alllivesmatter, PLOS ONE **13**, e0195644 (2018).

[17] K. Gothard, D. R. Dewhurst, J. R. Minot, J. L. Adams, C. M. Danforth, and P. S. Dodds, The incel lexicon: Deciphering the emergent cryptolect of a global misogynistic community, CoRR **abs/2105.12006** (2021), 2105.12006.

[18] H. H. Wu, R. J. Gallagher, T. Alshaabi, J. L. Adams, J. R. Minot, M. V. Arnold, B. F. Welles, R. Harp, P. S. Dodds, and C. M. Danforth, Say their names: Resurgence in the collective attention toward black victims of fatal police violence following the death of george floyd, PLOS ONE **18** (2021).

[19] United Nations Human Rights Office, Human rights violations and abuses related to the protests of july and august 2024 in bangladesh (2025).

[20] Ministry of Health, List of persons killed and injured in the mass uprising of students, As accessed on 24 January 2024 (2024).

[21] P. S. Dodds, K. D. Harris, I. M. Kloumann, C. A. Bliss, and C. M. Danforth, Temporal patterns of happiness and information in a global social network: Hedonometrics and twitter, PLoS ONE **6** (2011).
[22] P. S. Dodds, T. Alshaabi, M. I. D. Fudolig, J. W. Zimmerman, J. Lovato, S. Beaulieu, J. R. Minot, M. V. Arnold, A. J. Reagan, and C. M. Danforth, Ousiometrics and telegnomics: The essence of meaning conforms to a two-dimensional powerful-weak and dangerous-safe framework with diverse corpora presenting a safety bias, ArXiv **abs/2110.06847** (2021).
[23] D. M. Blei, A. Ng, and M. I. Jordan, Latent dirichlet allocation (2009).
[24] R. J. Gallagher, M. R. Frank, L. Mitchell, A. J. Schwartz, A. J. Reagan, C. M. Danforth, and P. S. Dodds, Generalized word shift graphs: a method for visualizing and explaining pairwise comparisons between texts, EPJ Data Science **10** (2020).
[25] T. Loughran and B. McDonald, When is a liability not a liability? textual analysis, dictionaries, and 10-ks, Journal of Finance **66**, 35 (2011).
[26] R. J. Gallagher, M. R. Frank, L. Mitchell, A. J. Schwartz, A. J. Reagan, C. M. Danforth, and P. S. Dodds, Generalized word shift graphs: a method for visualizing and explaining pairwise comparisons between texts, EPJ Data Science **10** (2020).
[27] P. Dodds, K. Harris, I. Kloumann, C. Bliss, and C. Danforth, Temporal patterns of happiness and information in a global social network: hedonometrics and twitter, PLoS ONE **6**, 26752 (2011).
[28] P. Dodds, J. Minot, M. Arnold, T. Alshaabi, J. Adams, D. Dewhurst, T. Gray, M. Frank, A. Reagan, and C. Danforth, Allotaxonometry and rank-turbulence divergence: a universal instrument for comparing complex systems, EPJ Data Sci. **12**, 37 (2023).
[29] R. J. Gallagher, A. J. Reagan, C. M. Danforth, and P. S. Dodds, Divergent discourse between protests and counter-protests: #blacklivesmatter and #alllivesmatter, PLoS ONE **13** (2016).

# Appendix

## A1. Example YouTube comments

| Bangla Comment | Translated Comment |
|---|---|
| স্বৈরাচার নিপাত যাক গণতন্ত্র মুক্তি পাক | Let the tyranny be overthrown and let the democracy be liberated |
| সব হত্যাকাণ্ডের সুস্পষ্ট তদন্ত এবং বিচার চাই | We want clear investigation and prosecution of all murders |
| শিক্ষার্থীদের উপর হামলার ঘটনায় তীব্র নিন্দা ও প্রতিবাদ জানাই | Strongly condemn and protest the attack on students. |
| সরকার যা শুরু করেছে একজন দেশদ্রোহী মতো কার্য শুরু করেছে মহান আল্লাহ গজব নাজিল হোক এই সরকারের উপর | What the government has started, it has started like a traitor, may Allah Almighty shower his wrath on this government |
| জীবনের অধিকার স্বাধীনতা না থাকলে কেবল রাষ্ট্রীয় স্বাধীনতার নামে বর্ডারের স্বাধীনতা- স্বাধীনতা নয়, ধোকার কারাগার। | If there is no right to life and freedom, border freedom is not only freedom in the name of state freedom, but a prison for fraud. |
| মানবিক গণতন্ত্র না থাকলে- সবার জীবনের নিরাপত্তা না থাকলে- শাসকগোষ্ঠীর অন্যায়ের প্রতিবাদের সুযোগ না থাকলে-রাষ্ট্র এক ধর্ম বা এক গোষ্ঠীর কুক্ষিগত জবরদখল হয়ে গেলে রাষ্ট্রের অভ্যন্তরীণ স্বাধীনতার মৃত্যু ঘটে এবং নাগরিকত্ব রুদ্ধ হয়ে জীবনের স্বাধীনতাও রুদ্ধ ও হরন হয়ে যায়। | If there is no human democracy - if there is no security for everyone's life - if there is no opportunity to protest against the injustice of the ruling group - if the state becomes a religion or a group's tyranny, then the internal freedom of the state will die and citizenship will be blocked and the freedom of life will also be blocked and taken away. |
| একক গোষ্ঠীর স্বৈরদস্যুতান্ত্রিক রাষ্ট্র স্বাধীন রাষ্ট্র নয় জীবনের স্বাধীনতাও নয়। | Not an independent state, not freedom of life. |
| সত্য ও মানবতার মুক্তির উৎস প্রাণাধিক প্রিয়নবী রাহমাতাল্লিল আলামীন সাল্লাল্লাহু আলাইহি ওয়াসাল্লাম প্রদত্ত সব মানুষের জীবন নিরাপত্তা অধিকার স্বাধীনতা মর্যাদা মালিকানা সেবা কল্যাণ ভিত্তিক ও একক গোষ্ঠীর স্বৈরদস্যুতামুক্ত অসাম্প্রদায়িক সর্বজনীন মানবতার রাষ্ট্র এবং মানবিক সাম্যের ভিত্তিতে বিশ্বসম্পদে সব মানুষের মালিকানা ভিত্তিক বিশ্বনাগরিকত্ব ভিত্তিক অখন্ড মানবতার অখন্ড দুনিয়া খেলাফতে ইনসানিয়াত ব্যতীত জীবনের স্বাধীনতা রাষ্ট্রের স্বাধীনতা ও বিজয় অলীক এবং মনস্তাত্ত্বিক ধোকা মাত্র। | The source of truth and liberation of humanity is the life of the beloved Prophet Rahmatallil Alamin sallallahu alaihi wasallam. Life security of all people, rights, freedom, dignity. Without Khilafat Insaniyat, the freedom of life, the freedom and victory of the state is just an illusion and psychological deception. |

Table A1. Example of Bangla comments with their English translations from 1000 videos during the movement period, July 16th, 2024 to August 5th, 2024, on YouTube related to the July Revolution and associated movement news videos.

## A2. Topic Model

We performed LDA topic modeling to analyze YouTube comments on movement-related videos in order to understand the topics people were discussing at that time. We selected 10 topics for the LDA model. Based on the associated words for each topic, we named each topic considering the main context of the movement, as the videos are exclusively related to the movement. Table A2 shows the topic names and their associated words.

| **Topic Name** | **Associated Words** |
|---|---|
| Media Discourse | time, say, ki, hands, ke, media, field, bhai, ar, indian, hand, open, bichar, chai, successful, face, nai, amra, firing, expatriates, better, gone, hok, action, boo, suffering, information, jonno, agitators, koto |
| Civic Unity | come, right, people, day, country, love, sir, common, quot, demand, ahead, wants, world, win, long, best, saying, protect, vai, brothers, use, anti, lost, point, man, national, means, forward, lives, 2024 |
| Hasina Politics | bangladesh, hasina, country, inshallah, sheikh, people, like, victory, save, salute, truth, stand, india, life, away, boycott, follow, video, need, arrested, fake, kill, brothers, accept, years, bad, run, said, shame, seen |
| Justice Demand | br, government, want, justice, children, nation, resignation, army, god, fall, teachers, die, given, family, bangla, loading, forces, dictatorship, war, abu, life, rights, db, state, free, end, saeed, ai, home, injustice |
| Digital Protest | https, police, whatsapp, channel, com, 0029vakgkm5ldqeemosdqc0f, href, join, jamunatelevisionofficial, er, watch, attack, telegram, good, force, terrorists, sob, ra, desh, korte, really, strongly, job, kore, era, valo, condemn, far, haroon, ora |
| Party Conflict | league, country, people, awami, students, chhatra, freedom, jamaat, bnp, minister, stop, drama, money, power, law, make, politics, prime, stay, shibir, way, party, doing, leaders, independent, days, fighters, new, terrorist, history |
| Prayerful Resistance | allah, na, brothers, support, joy, death, brother, amen, home, ka, ameen, help, kore, journalist, hobe, dear, little, ta, der, kora, grant, kotha, soon, lot, getting, kono, jomuna, ei, koro, mukti |
| News Debate | tv, jamuna, people, don, thanks, let, br, news, know, yamuna, television, does, resign, judge, understand, like, want, killing, words, live, respect, father, think, court, public, brother, coming, talk, tell, common |
| Global Outreach | youtube, href, www, com, http, results, searchquery, e0, br, a6, thank, a7, savebangladeshistudents, 23savebangladeshistudents, bengal, stepdownhasina, a8, 23stepdownhasina, 23, amp, 23unitednations, unitednations, 8b, 95, newyorkpost, thewallstreetjournal, 23thewallstreetjournal, 23cnn, cnn, projectnightfall |
| Student Violence | students, movement, police, student, alhamdulillah, today, did, blood, killed, quota, sisters, country, hall, streets, shot, leave, continue, going, died, brother, situation, happen, mother, red, chest, demands, gave, murder, request, war |

Table A2. Top 10 topics extracted from LDA topic modeling on YouTube Comments Related to News, with associated words for each topic.

As a few topics are closely related, we merged these topics and renamed them with new names. After this, the optimum number of topics becomes 6, as Global Outreach and Student Violence remain unmerged. Table A3 shows the merged topics with new names.

| **Merged Topics** | **New Topic Name** |
| --- | --- |
| Hasina Politics + Party Conflict | Political Conflict |
| Civic Unity + Prayerful Resistance | Social Resistance |
| Media Discourse + News Debate | Media Discourse (content merged, name retained) |
| Digital Protest + Justice Demand | Digital Movement |
| Global Outreach | Global Outreach |
| Student Violence | Student Violence |

Table A3. Closely related topics merged together and renamed new topic.